\documentclass[draft]{agujournal2019}
\usepackage{url} 
\usepackage{lineno}
\usepackage[inline]{trackchanges} 
\usepackage{soul}
\usepackage[colorinlistoftodos,prependcaption,textsize=tiny]{todonotes}
\linenumbers
\draftfalse

\journalname{Journal of Advances in Modeling Earth Systems}

\nolinenumbers
\begin{document}

%
%


\title{An Ensemble Machine Learning Approach for Tropical Cyclone Detection Using ERA5 Reanalysis Data}

%
%




\authors{Gabriele Accarino\affil{1}, Davide Donno\affil{1}, Francesco Immorlano\affil{1,2}, Donatello Elia\affil{1}, Giovanni Aloisio\affil{1,2}}

\affiliation{1}{Advanced Scientific Computing Division, Centro Euro-Mediterraneo sui Cambiamenti Climatici, Lecce, Italy}
\affiliation{2}{Department of Innovation Engineering, University of Salento, Lecce, Italy}





\correspondingauthor{Gabriele Accarino}{gabriele.accarino@cmcc.it}



\begin{keypoints}
\item A TC detection approach based on Machine Learning (ML) is proposed using ERA5 reanalysis and IBTrACS records.
\item The approach was able to accurately detect lower TC categories than those used for training the models.
\item The ensemble approach was able to further improve TC localization performance with respect to single model TC center estimates.
\end{keypoints}

%
%

%
%


\begin{abstract}
Tropical Cyclones (TCs) are counted among the most destructive phenomena that can be found in nature. Every year, globally an average of 90 TCs occur over tropical waters, and global warming is making them stronger, larger and more destructive. The accurate detection and tracking of such phenomena have become a relevant and interesting area of research in weather and climate science. Traditionally, TCs have been identified in large climate datasets through the use of deterministic tracking schemes that rely on subjective thresholds. Machine Learning (ML) models can complement deterministic approaches due to their ability to capture the mapping between the input climatic drivers and the geographical position of the TC center from the available data. This study presents a ML ensemble approach for locating TC center coordinates, embedding both TC classification and localization in a single end-to-end learning task. The ensemble combines TC center estimates of different ML models that agree about the presence of a TC in input data. ERA5 reanalysis were used for model training and testing jointly with the International Best Track Archive for Climate Stewardship records. Results showed that the ML approach is well-suited for TC detection providing good generalization capabilities on out of sample data. In particular, it was able to accurately detect lower TC categories than those used for training the models. On top of this, the ensemble approach was able to further improve TC localization performance with respect to single model TC center estimates, demonstrating the good capabilities of the proposed approach. 
\end{abstract}

\section*{Plain Language Summary}
Every year an average of 90 Tropical Cyclones occur globally and this number is expected to increase due to global warming, which is also increasing the frequency and the intensity of such extremes. The detection and tracking of Tropical Cyclones has been traditionally addressed through the use of deterministic tracking schemes. Machine Learning  can complement traditional schemes by providing an end-to-end approach to learn the relationships between the climatic drivers and the cyclone center position, directly from the available data. In the present study, an ensemble approach that locates the center coordinates is introduced. Basically, the idea is to rely on several Machine Learning models accomplishing the same task - each with different training configurations - to integrate their results and achieve higher localization accuracy than a single Machine Learning model. The climate variables, used as predictor for training and testing of the models, were gathered from ERA5 reanalysis, while the historical Tropical Cyclones center positions, used as target, were retrieved from the International Best Track Archive for Climate Stewardship dataset. The results showed the effectiveness of the proposed approach against the use of a single Machine Learning model.


%
%

%


%
%
%
%

\section{Introduction}
Tropical Cyclones (TCs), also known as hurricanes or typhoons, are counted among the most fascinating and destructive phenomena that can be found in nature \cite{kerry2003}. Several conditions are at the basis of TC formation. As described by \citeA{riehl2004} and \citeA{dunn1951}, the process is triggered by warm ocean waters’ condensation that provides most of the energy to the system. A sufficient distance of the TC from the equator allows the Coriolis force to take effect, leading to  the typical spinning motion. In addition, stored heat energy is released by condensation, warming up the air and lowering the pressure. Besides the aforementioned conditions, the cyclone center (i.e., its eye) is typically located in a low-pressure region, surrounded by strong winds and deep cumulonimbus. As the TC travels, it becomes an auto-sufficient system that continuously gathers energy from the ocean. If the TC shifts on land (i.e., the so-called landfall), the TC loses the energy drawn by warm water’s condensation, thus leading to its rapid dissipation \cite{kepert2010, ruttgers2019, metoffice2023}.

The geographical areas in which the formation of TCs is supported are called \textit{cyclone formation basins}. There are seven basins around the world, each with a specific water depth and sea surface temperature: the main consequence is the difference in the number of TCs per year and the season in which they develop \cite{roy_kovordanyi2012}. Every year, globally an average of 90 TCs occur over tropical waters \cite{emanuel_nolan2004} and global warming is making them stronger, larger and more destructive, as recognized by \citeA{elsner2008, mendelsohn2012, sun2017}. As reported by the World Meteorological Organization (WMO), $1,942$ disasters have been attributed to TCs, which caused US \$ $1,407.6$ billion in economic losses and almost 8 million killed people over the past 50 years \cite{wmo2023}, thus making TCs impact quite significant on different sectors, such as infrastructures, economy, human health, social unrest.

The accurate detection and tracking of such phenomena have become a relevant and interesting area of research in weather and climate science \cite{dabhade2021, scoccimarro2014}. Traditionally, TCs have been identified in large climate datasets through the use of deterministic tracking schemes, also known as TCs trackers \cite{horn2014}. The latter are algorithms capable of identifying — by means of thresholds applied on variables significant to the cyclogenesis —  patterns related to a warm core in gridded datasets and connecting them along the TC trajectory \cite{bourdin2022}. Depending on the particular variables involved in the tracking process, two main categories of schemes exist: physics-based (see \citeA{camargo2002, zhao2009, zarzycki2017, murakami2014, horn2014, chauvin2006}) and dynamics-based that include the TRACK method \cite{hodges2017, strachan2013} and the Okubo-Weiss-Zeta (OWZ) algorithm \cite{tory2013}.

The aforementioned thresholds are set by the author of the scheme, therefore they are subjective and mainly rely on the human expertise about the phenomena under investigation \cite{dabhade2021, enz2022}. Moreover, thresholds may depend on the particular geographical region of study and the related formation basin, as well as on the TC categories \cite{bloemendaal2021, befort2020}. However, manual threshold tuning may lead to subjective bias, as well as to the potential inability of generalizing the proposed approach to other situations or domains.

The state of different climate variables, which the tracking schemes are applied on, is simulated by physics-based Earth System Models (ESMs) that provide large amounts of data at different spatio-temporal resolutions. In addition to ESM data, ground-based in situ observations and satellite retrievals contribute to further increase the data volume. Such large-scale data introduces issues in terms of how scientific data can be effectively managed and processed to make the best out of it \cite{gray2005}. Indeed, climate scientists, meteorological agencies and policy decision makers need to process and extract meaningful information from these huge datasets in a cost-effective manner and in a reasonable amount of time \cite{sebestien2021}. In this context, High Performance Data analytics systems can address some of the issues and provide support for descriptive/statistical analysis from this large-scale data \cite{elia2021}. Nevertheless, in the last few years Machine Learning (ML) and Deep Learning (DL) algorithms became popular as data-driven paradigms for supporting feature extraction from the vast amounts of scientific data currently available \cite{hey2020}. ML and DL algorithms can, actually, go beyond what can be extracted with traditional descriptive and deterministic methodologies. In particular, for the phenomena targeted by this study, ML models are well-suited since they can capture the mapping between the input climatic drivers and the geographical position of the TC center from the available data, without the need of subjective thresholds. As a consequence, such models can complement deterministic tracking schemes for the TC detection task. 

To this extent, several research efforts can be found in the scientific literature towards the development of cutting-edge TC detection approaches beyond the existing deterministic tracking schemes. For example, many studies focused on the use of satellite data and DL approaches for accurately locating the TC eye \cite{carmo2021}. 

Several works, such as \citeA{lam2023, pang2021, snehlata2020, haque2022}, framed the identification of the TC center as an object detection task for which the You Only Look Once (YOLO) v3 DL object detection model was adopted. Similarly, a DL-based object detection approach was proposed in \citeA{wang2020} with the aim of retrieving the TC center through segmentation, edge detection, circle fitting, and comprehensive decision of satellite IR images. Segmentation of the shape and size of the detected TC in high-resolution satellite images was also provided by \citeA{nair2022}. To this extent, a pipeline consisting of a Mask Region-Convolutional Neural Network (R-CNN) detector, a wind speed filter and a CNN classifier was adopted to accurately detect TCs. In \citeA{xie2022}, a Feature Pyramid Network (FPN) was proposed as a feature extractor and region proposal network that searches for the potential areas of cyclones along with a \textit{Faster Region-based CNN} (Faster R-CNN) to calibrate the locations of TC regions. Faster R-CNN was also used by \citeA{xie2020} to classify the presence of TCs in Mean Wind Field-Advanced Scatterometer (MWF-ASCAT) satellite data. The authors of \citeA{kim2019b} exploited a Convolutional LSTM (ConvLSTM) to detect, track and predict hurricane trajectories on Community Atmospheric Model v5 (CAM5) simulation data. With the aim of capturing both temporal dynamics and spatial distribution, trajectories were modeled as time-sequential density maps. The detection of tropical and extratropical cyclones was addressed as an image segmentation task by \citeA{kumlerbonfanti2020}. They used Global Forecasting System (GFS) and Geostationary Operational Environmental Satellite (GOES) to compare four state-of-the-art U-Net-based models designed for the detection task. In \citeA{carmo2021}, data from the Sentinel-1 C-band SAR satellite were used to provide a DL-based detector of the TC center, also providing estimates of the related category according to sea surface wind and rain-related topological patterns. Authors further provided explainability through the analysis of key patterns highlighted by the Gradient-based Class Activation Map method. \citeA{kim2019a} used eight predictors gathered from the WindSat satellite to frame a TC detection task. Then, they compared the detection skills of three ML algorithms, namely Decision Trees (DT), Random Forest (RF), and Support Vector Machines (SVM) and a model based on Linear Discriminant Analysis (LDA).

A TC detection approach based on ML is proposed in this work and applied on the joint North Pacific and Atlantic TC formation basins. Although the detection task is similar to other studies from the state of the art, there are some important differences in the algorithmic approach used. From a methodological perspective, a ML ensemble approach is proposed to accurately locate the TC center coordinates. Exploiting a single ML model for locating the TC center would have resulted in unreliable results because of the inherent complexity of the TC detection task. The ensemble, instead, allows combining TC center estimates of different ML models that are in agreement about the presence of a TC in input data. In this way, each model can learn different spatial characteristics of the TC structure and the ensemble allows providing more accurate TC center estimates. Additionally, the ML setup used in this study allows embedding both TC classification and localization in a single end-to-end learning task.

Moreover, in contrast to other studies, a total of six ERA5 reanalysis TCs predictors (i.e., mean sea level pressure, 10m wind gust since previous post-processing, instantaneous 10m wind gust, relative vorticity at 850 mb and temperature at 300 and 500 mb) were used in place of satellite data. Reanalysis data combines model simulations and observations to provide the best representation of climate variables in the past \cite{ecmwf2020}. Accurate TC centers geographical coordinates were retrieved from the International Best Track Archive for Climate Stewardship (IBTrACS) dataset, the most complete global collection of historical TC occurrences \cite{ibtracswebsite}.

The rest of the paper is organized as follows: Section \ref{sec2} describes data sources and the processing steps required to build a suitable dataset for ML training. Moreover, the experimental setup is described, along with models architectures and the ensemble procedure. Section \ref{sec3} presents the results on the test set, focusing on some meaningful test cases for which the performance of models’ ensemble is compared against single model ones. Section \ref{sec4} discusses the obtained results, highlighting strengths and limitations of the proposed approach, also drawing the main conclusions from this work and points out some relevant future activities.

\section{Materials and Methods}\label{sec2}
    \subsection{Data Sources}
    This subsection provides a description of the two data sources used to build the dataset for the ML setup.
    
        \subsubsection{The International Best Tracks Archive for Climate Stewardship}
        The International Best Track Archive for Climate Stewardship (IBTrACS) presented by \citeA{knapp2010} is an institutional, open access and centralized archive that provides the most complete set of historical TC best track data at a global level. It integrates historical records with observations retrieved from 12 different meteorological agencies. The main aim of IBTrACS is fostering research in the context of such events by keeping track of their geographical position, frequency and intensity worldwide. IBTrACS reports global TCs occurrences at $0.1^{\circ}$ ($\sim 10 km$) of spatial resolution from 1841 to present with a 3-hourly temporal frequency. However, in this study, only TC records between 1980 and 2019 were selected from IBTrACS v4 \cite{knapp2018} at a temporal frequency of 6 hours. Although IBTrACS provides TC records from 1841 to date, 1980 is considered the beginning of the Modern Era, characterized by the extensive use of geostationary satellite imagery at a global scale. On the other hand, more recent TC information are subject to frequent reanalysis by the different meteorological agencies contributing to IBTrACS, and, for these reasons, the TC selection was limited to 1980-2019. Furthermore, 6-hourly data provides additional information about the TC characteristic, such as the Maximum Sustained Wind (MSW), contrary to 3-hourly ones \cite{knapp2010}. 
        
        Concerning the geographical domain, this study mainly targets the North Pacific formation basin which is widely recognized as a particularly active region where most TCs occur every year \cite{roy_kovordanyi2012}. Since a substantial number of TC events cross both the North Pacific and North Atlantic regions, thus reaching up to $320 ^{\circ}E$ of longitude, the final domain of interest is $100-320 ^{\circ}E$, $0-70 ^{\circ}N$ (i.e., joint North Atlantic and North Pacific). 
        
        \subsubsection{ERA5 Reanalysis}
        Climate variables that are the main drivers of TCs, thus contributing to the formation and sustainment during their lifetime, were retrieved from the Copernicus Climate Change Service (C3S) ERA5 reanalysis datasets. ERA5 reanalysis combines global numerical weather predictions with newly available observations in an optimal way to produce consistent estimates of the state of the atmosphere \cite{reanalysis2020}. In this study, mean sea level pressure $[Pa]$ (\textit{msl}), 10m wind gust since previous post-processing $[ms^{-1}]$ (\textit{fg10}) and the instantaneous 10m wind gust $[ms^{-1}]$ (\textit{i10fg}) were gathered from the ERA5 reanalysis on single levels \cite{hersbach2023a}, whereas the relative vorticity at 850 mb $[s^{-1}]$ (\textit{vo850}) and the temperature at 300 and 500 mb $[K]$ (\textit{t300} and \textit{t500}, respectively) were collected from the ERA5 reanalysis on pressure levels dataset \cite{hersbach2023b}. Each of the aforementioned climatic variables  was provided on a regular grid of  $0.25^{\circ} \times 0.25^{\circ}$ ($\sim 27 km \times 27 km$) of spatial resolution, targeting the geographical domain previously described and it was managed as a 2-dimensional map of size $280 \times 880$ pixels. Moreover, data was collected with a 6-hourly temporal resolution (i.e., $00.00$, $06.00$, $12.00$ and $18.00$ time steps) for the period 1980-2019, thus matching TCs records selected from IBTrACS, except for \textit{fg10} that was originally collected with a hourly temporal resolution. In particular, the ERA5 \textit{fg10} variable reports the maximum of the wind gust in the preceding hour. Therefore, to match the 6-hourly temporal resolution of this study, for each time step, the maximum over the previous 6 hours was computed.

    \subsection{Data Processing}
        \subsubsection{IBTrACS filtering and selection}
        Starting from trajectories belonging to the joint North Atlantic and North Pacific geographical domain ($100-320 ^{\circ}E$, $0-70 ^{\circ}N$), only IBTrACS records with \textit{track\_type} field flagged as \textit{main} were considered. Therefore, \textit{provisional}, \textit{spur} and \textit{provisional-spur} tracks were implicitly discarded as they are characterized by a higher level of uncertainty \cite{ibtracs_doc2019}. It is noteworthy that data from recent years are typically provided as \textit{provisional} or \textit{spur}, meaning that their corresponding values have not been reanalyzed yet and therefore are of lower quality compared to \textit{main} tracks. This can happen because some variables — such as the intensity, position and storm categories — are subject to change according to posterior reanalysis by the meteorological agencies. Moreover, uncertainties in the observing system may result in contradictory opinions by different agencies about the storm location, leading to \textit{spur} tracks. This is mainly due to difficulties in localizing the center of circulation or in the case of storms merging (i.e., Fujiwhara effect) \cite{ibtracs_doc2019}. As an additional selection step, tracks were filtered out based on the nature field, specifically discarding those trajectories marked as: \textit{(i)} \textit{Not Reported} (\textit{NR}) whose nature is unknown, \textit{(ii)} \textit{Disturbance Storms} (\textit{DS}) that correspond to not-well-formed storms characterized by a MSW less than 34 knots, and \textit{(iii)} \textit{Mixture} (\textit{MX}) that correspond to tracks that received contradicting reports about the nature of the observing system from different agencies. At the end of this filtering and selection process, only \textit{Tropical Storms} (\textit{TS}),\textit{Extra Tropical} (\textit{ET}) and \textit{Subtropical Storms} (\textit{SS}) main tracks were considered at a 6-hourly temporal resolution.
        
\begin{sidewaysfigure}
        \noindent\includegraphics[width=\textwidth]{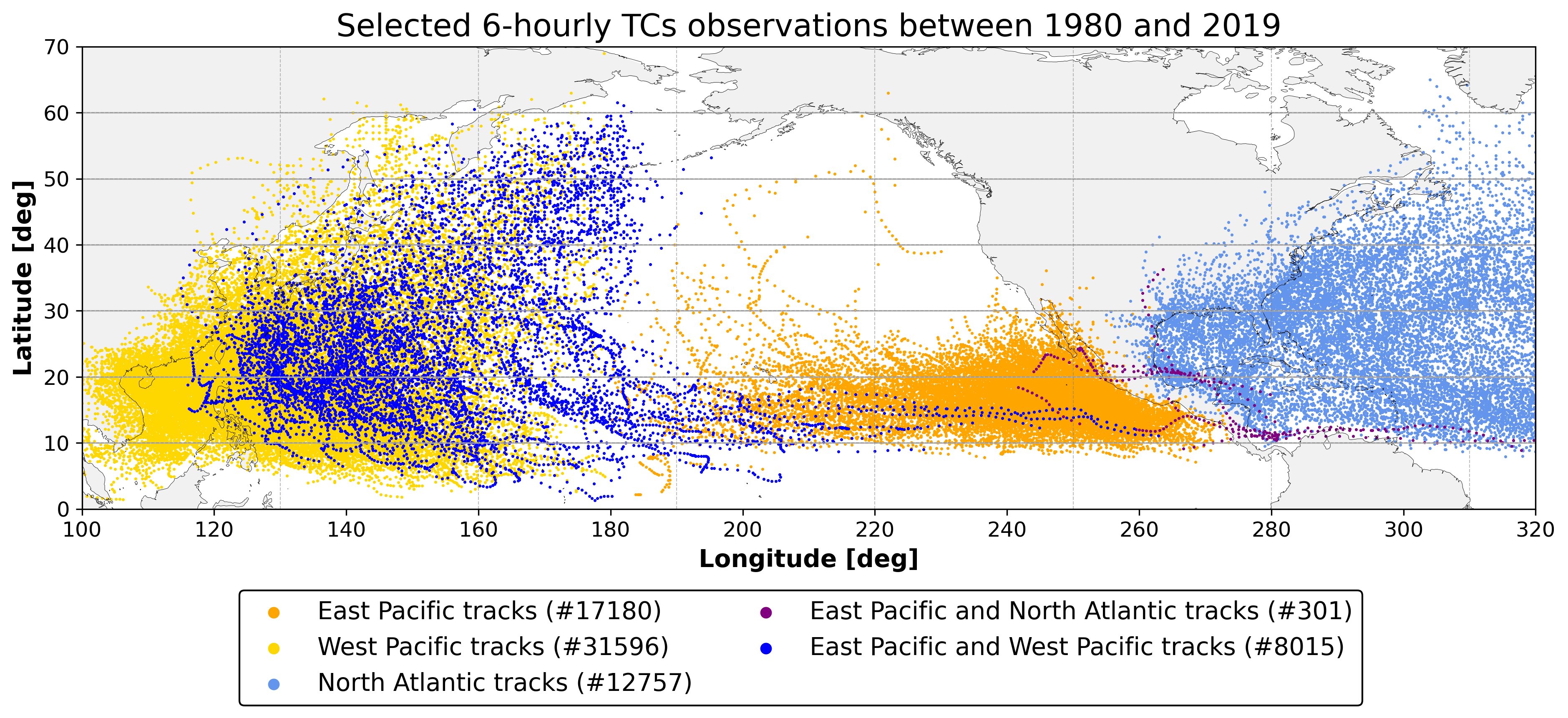}
        \caption{Visualization of 6-hourly TCs location within the 1980-2019 period in the joint North Atlantic and North Pacific geographical domain ($100-320 ^{\circ}E$, $0-70 ^{\circ}N$). TCs in each sub-basin are highlighted by a different color, along with the relative number of occurrences. Only IBTrACS records whose nature is TS, SS and ET are shown in the picture.}
        \label{fig:selected_tracks}
        \end{sidewaysfigure}

        Figure \ref{fig:selected_tracks} shows the selected TC records occurring in the domain of study for the considered period. TCs tracks were further divided into non-overlapping groups according to the route taken during their lifecycle, reporting also the number of occurrences in each group. The trajectory of TCs in the North Atlantic (light blue), West Pacific (yellow) and East Pacific (orange) remain confined to such basins (i.e., they originate and dissipate in the same basin), whereas East and West Pacific tracks (blue) as well as East Pacific and North Atlantic ones (purple) cross different basins during their lifecycle. 

        \subsubsection{Patches generation and labeling}\label{sec222}
        For each of the six climatic drivers, ERA5 maps (of dimension $280 \times 880$ pixels) were evenly tiled into $7 \times 22$ non-overlapping patches of size $40 \times 40$ pixels each. The TC eye can occur in every pixel of the patch, not necessarily in its center, thus non cyclone-centric patches were generated. Then, drivers were stacked together, resulting in data of dimension $40 \times 40 \times 6$. In order to associate patches containing a TC (from now on referred to as \textit{positive patches}) with its center position (i.e., the TC eye), the latitude and longitude geographical coordinates extracted from IBTrACS were rounded to match the resolution of the ERA5 grid ($0.25^{\circ} \times 0.25^{\circ}$). Subsequently, rounded coordinates were further converted into local-patch positions in terms of row-column index pairs (considering the $40 \times 40$ patch as a matrix).
        
        Different TC phenomena may simultaneously occur in the domain of interest at a particular time, thus multiple \textit{positive patches} can be retrieved from a single ERA5 map. The patches that do not contain a TC (from now on referred to as \textit{negative patches}) were labeled with a negative row-column coordinate (i.e., $[-1,-1]$), indicating the TC absence. In this way, retrieved local patch row-column pairs were used as the target of the detection task.

    \subsection{Experimental Setup}
        \subsubsection{Dataset creation}\label{sec231}
        Selected patches in the considered period (1980–2019) were split into training and test sets as follows: patches belonging to two consecutive years for each decade were selected for testing (1983, 1984, 1993, 1994, 2003, 2004, 2013, 2014, respectively), whereas patches in the remaining years were used for training. In this way, the training set comprises patches that span the whole time period, enabling ML models to capture and learn potential climate change patterns that may affect the input atmospheric drivers \cite{essentialclimatevariables}.
        
\begin{sidewaysfigure}
        \noindent\includegraphics[width=0.6\textwidth]{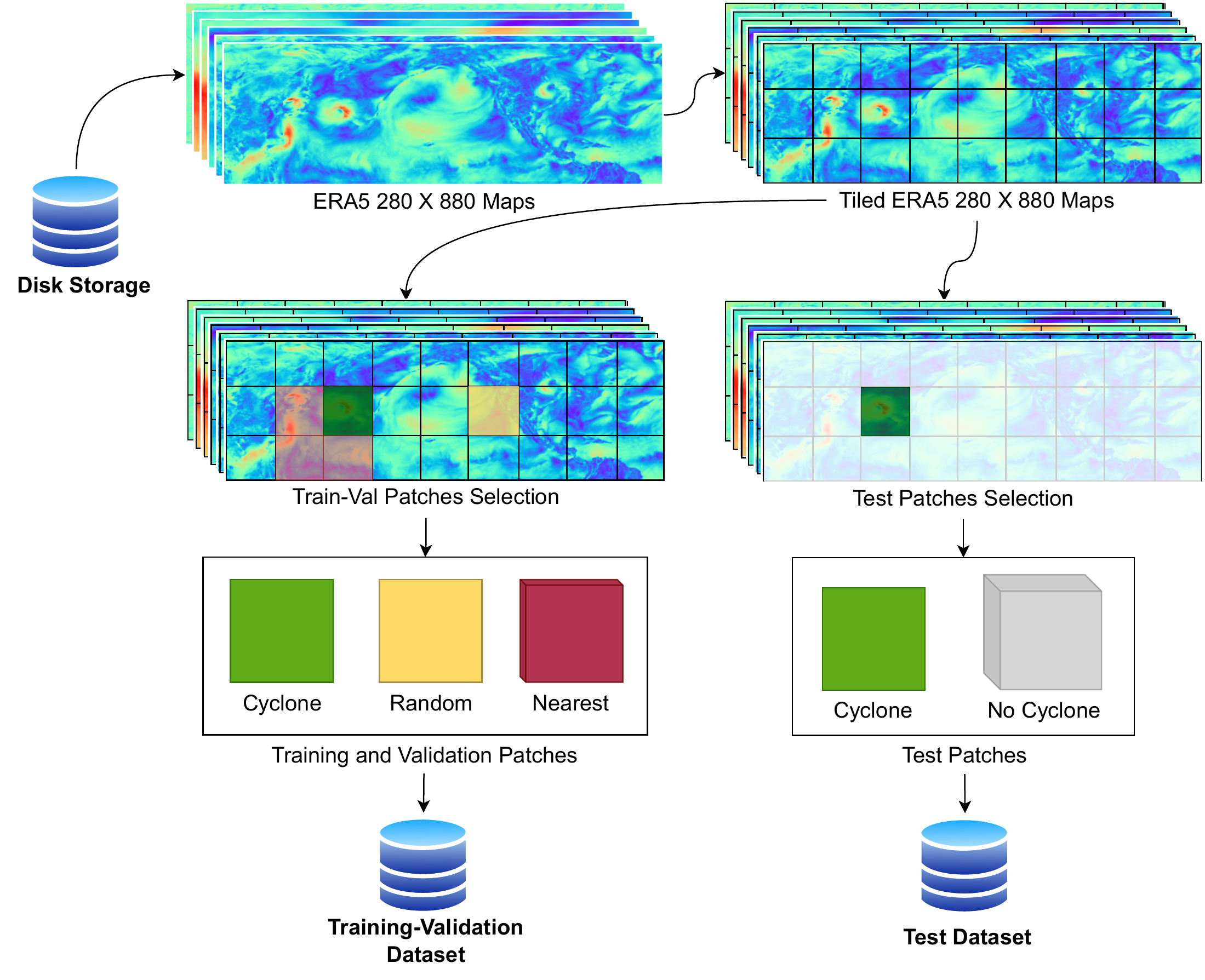}
        \caption{Overview of the dataset building pipeline. ERA5 maps are tiled into non-overlapping patches of size $40 \times 40$. To create the training and validation subsets, for each tiled ERA5 map, the patch containing the TC is considered along with nearest and random patches, discarding the remaining ones. Concerning the test subset, all the patches are considered.}
        \label{fig:patches_pipe}
        \end{sidewaysfigure}

        In order to build the training set, \textit{negative patches} were carefully selected to enhance the variance of the dataset, as well as to improve the predictive skills of ML models. Among the edge patches surrounding a \textit{positive} one, the three corner patches closest to the storm center were considered as \textit{negative} (referred to as \textit{nearest patches}, see purple patches in Figure \ref{fig:patches_pipe}). Despite nearest patches are labeled as \textit{negative}, they may contain residual structures (e.g., spiral wind gust tails, minimum regions of mean sea level pressure, etc.) of the TC located in the central patch. Therefore, including such patches can benefit model training. 
        
        Additionally, for each \textit{positive patch} a further \textit{negative sample} was randomly selected among the $7 \times 22$ patches of the map excluding the edge ones previously mentioned and thus ensuring that no major TC phenomena occur in the randomly selected patch.
        By construction, the training set is imbalanced towards \textit{negative samples} (i.e., $55,639$ \textit{positive patches} and  $212,679$ \textit{negative} ones, yielding $20\%$ of samples containing a TC). To address this imbalance ratio, as well as to increase the variance of the training set, a selective data augmentation procedure was used. For each positive sample, three transformations were considered: \textit{left-right flip}, \textit{up-down flip} and \textit{$180^{\circ}$ rotation} \cite{shorten_khoshgoftaar2019}. 
        
        Conversely, all the $7 \times 22$ patches belonging to each map of the test set years were selected to assess the actual performance of ML models on out-of-sample data. The resulting test set consists of $967,513$ negative patches and $14,149$ positive ones, ending up in a strongly imbalanced test set (i.e., only $1.46\%$ of positive samples).

        \subsubsection{Neural Network Architectures}
        For each input patch that comprises the six climatic drivers, the proposed TC detection task aims at predicting the local row-column coordinates corresponding to the TC center, if present. From a logical point of view, the detection task can be split into two consecutive subtasks, i.e., \textit{classification} and \textit{localization}. Classification consists in determining the presence or absence of TCs within the input patches, whereas the localization subtask concerns the  prediction of the TC center coordinates, if present.
        
        To this extent, several VGG-like Neural Network (NN) architectures \cite{simonyan2014} were developed and trained, each differing in terms of number of layers, filters and kernel sizes. 
        
\begin{sidewaysfigure}
        \noindent\includegraphics[width=\textwidth]{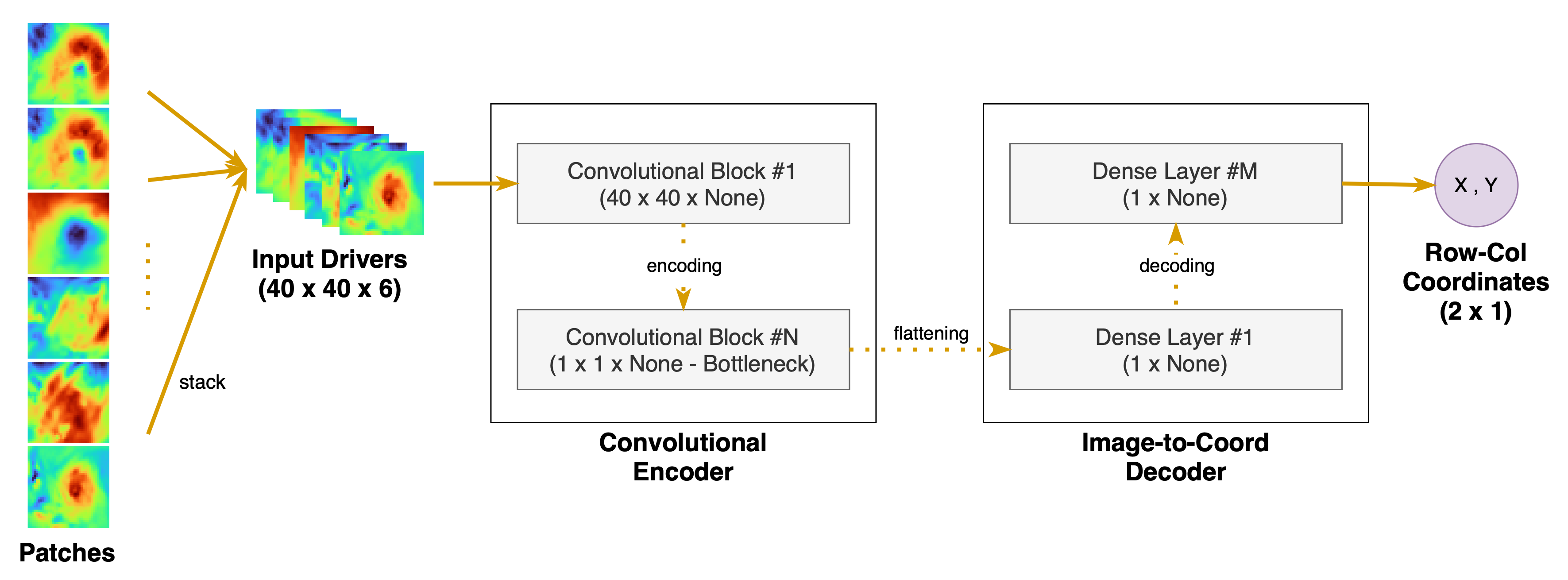}
        \caption{Basic representation of VGG-based models configured to address the TC detection task. Patches related to the six input drivers are stacked together and the local row-column coordinates of the TC center are used as the target. Models differ for complexity as well as different hyperparameters configuration.}
        \label{fig:models_arch}
        \end{sidewaysfigure}

        Figure \ref{fig:models_arch} depicts the overall architecture. Input patches are processed by a series of convolutional and max-pooling layers that encode the input volume, thus progressively decreasing height and width dimensions, while, at the same time, increasing the depth of the activation volume. The encoder ends with a bottleneck layer (i.e., the N-th convolutional block in Figure \ref{fig:models_arch}) that squeezes the processed information resulting in a lower dimensional representation of the patch content \cite{bottleneck}. In the decoding stage, bottleneck output is flattened and processed through a series of dense layers that gradually reconstructs the information and links it with the target row-column coordinates of the TC center within the patch. In this sense the VGG-like network is trained to learn the mapping between input climatic drivers content and the two output coordinates.

        A total of four different model configurations have been assessed for the detection task, called VGG V1, VGG V2 and VGG V3 that differ for complexity, as well as VGG V4 which was obtained by replacing each basic convolutional layer with a customized one: Conv + Gaussian Noise (optional) + Batch Normalization (optional) + Dropout (optional) + Leaky ReLU \cite{maas2013}. Optional hyperparameters are selectively activated on each layer. Additionally, the first two convolutional layers of VGG V4 are variable in the kernel size, enabling the model to capture wider spatial features, thus including the TC eye and its surrounding structure, as well.

        \subsubsection{Training and Validation}
        For each model configuration, $25\%$ of the training patches were used for validation purposes, whereas the remaining for the actual training. The six input drivers were normalized in the $[0,1]$ range and augmented according to the data augmentation procedure presented in Section \ref{sec231}. The aforementioned models were trained for 500 epochs with a batch size of $8,192$ patches, using the Adam optimizer \cite{kingma2014} with a learning rate of $1e^{-4}$.
        
        Two different loss functions were implemented and used. The first one is the Mean Absolute Error (MAE) between real and predicted coordinates. The second one is a custom loss defined by the authors for this study, called \textit{Cyclone Classification Localization} (CCL) loss, which is a linear combination of the MAE, the Binary Cross Entropy (BCE) loss and the Euclidean distance between real and predicted coordinates (L2). CCL tries to contemporarily pursue two goals: \textit{(i)} minimize the classification error (through BCE) and \textit{(ii)} minimize the localization error (through MAE and L2 terms). Based on different hyperparameters configuration, a total of 13 models were trained.

        \subsubsection{Test Metrics}\label{sec234}
        The test set was used to evaluate the generalization capabilities of the trained ML models on out-of-sample data. To this extent, different evaluation metrics were computed according to the classification and localization subtasks:

        \begin{itemize}
            \item Classification metrics: False Positives (FP), True Positives (TP), False Negatives (FN) and True Negatives (TN) rates were computed, along with precision and recall. It is worth noting that in this context a FP occurs when the ML model incorrectly classifies a patch as containing a TC although it is not present in reality.
            \item Localization metrics: the Euclidean distance between the real and predicted TC center coordinates was computed only for those patches correctly classified as positive (TP) by the ML model.
        \end{itemize}
        
        \subsubsection{Consensus and Models Ensemble}\label{sec235}
\begin{sidewaysfigure}
        \noindent\includegraphics[width=\textwidth]{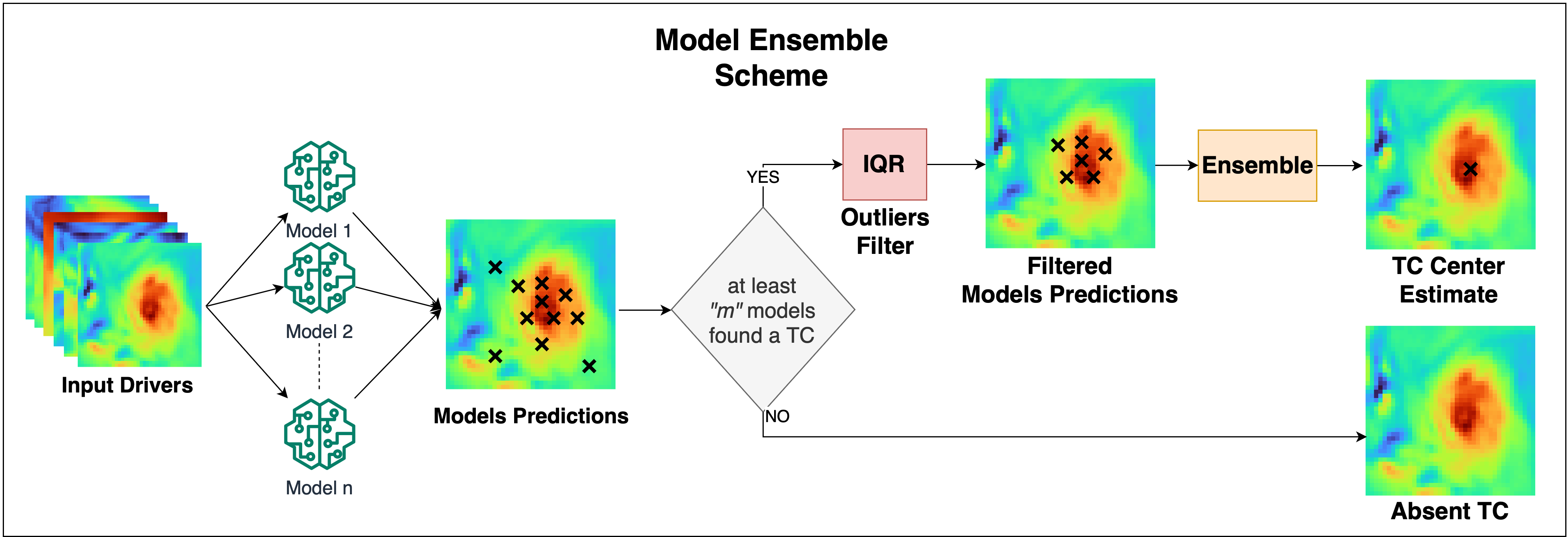}
        \caption{Diagram representing the models ensemble approach. All the \textit{n} pre-trained models are fed with the same patches, yielding \textit{n} row-column couples. If less than m (given) models detect the presence and localize the TC in the patch, the TC is considered as absent and the patch is labeled with negative coordinates. Otherwise, the IQR algorithm is applied on the output of the models detecting the TC; the final estimated location of the TC center is computed by averaging the values of the predictions not filtered by IQR.
        }
        \label{fig:ensemble_scheme}
        \end{sidewaysfigure}

        Since the 13 models are trained with a different set of hyperparameters and/or layers configuration, each of them learns different characteristics and high-level features in the training set patches. Therefore, an ensemble approach \cite{ganaie2022} has been assessed in this study to combine the predictions made by different models with the aim of improving the overall accuracy skills (see Figure \ref{fig:ensemble_scheme}). 

        As depicted in Figure \ref{fig:ensemble_scheme}, for each patch of the test set, the approach consists in evaluating first how many models agree in classifying it as positive. The number of models that should be in agreement — \textit{m} in Figure \ref{fig:ensemble_scheme} — represents an additional hyperparameter that indicates the minimum level of consensus required to consider the patch as likely containing a TC. 
        
        After a trial and error procedure on the validation set, this parameter was set to 7 in order to reach a trade-off between FP and FN rates on the test set, given that a lower number of FNs is preferable for the task of predicting the occurrence of such Extreme Weather Events (EWEs). Each of the 13 models can potentially provide very different estimates about the location of the TC center for the same input patch. Therefore, the Interquartile Range (IQR) method was adopted as a further filtering step to keep only the estimates closer to their median value. In particular, the method consists in considering as outliers those TC center estimates (\textit{x}) that satisfy the following inequality:

        \begin{linenomath*}
        \begin{equation}
        x <Q_1 - 1.5*IQR \lor x > Q_3 + 1.5*IQR
        \end{equation}
        \end{linenomath*}

        Indeed, the IQR is computed as the difference between the third ($Q_3$) and the first ($Q_1$) quartile, providing information about the spread of the data around the median value. Finally, the localization of the TC center is performed as the ensemble average of the row-column estimates of inliers.

\section{Results}\label{sec3}
    Table \ref{table:1} summarizes the averaged results produced by the 13 models on the test set, according to the evaluation metrics presented in Section \ref{sec234}. These 13 models are involved in the ensemble approach described in Section \ref{sec235}.

    \begin{sidewaystable}
    \caption{Average metrics over the test set for each of the 13 models. Ensemble performance on both imbalanced and randomized balanced test sets was also reported for comparison in the last two rows.}
    \centering
    \begin{tabular}{p{0.5cm} p{4cm} p{1.2cm} p{1.2cm} p{2cm} p{1.5cm} p{1.5cm} p{1.5cm} p{1.5cm} p{2.2cm} p{1.8cm}}
    \hline
    \# & Model type & Loss & Kernel size & Euclidean distance on TPs (km) & FP on NoCyclone (\%) & TP on Cyclone (\%) & FN on Cyclone (\%) & TN on NoCyclone (\%) & Precision (\%) & Recall (\%) \\
    \hline
    1 & VGG V1 & mae & 3 & \textbf{128.84} & 6.02 & 89.67 & 10.33 & 93.98 & 17.88 & 89.67 \\
    2 & VGG V2 & mae & 3 & 145.23 & 11.12 & 91.74 & 8.26 & 88.88 & 10.77 & 91.74 \\
    3 & VGG V3 & mae & 3 & 151.61 & 12.71 & 91.35 & 8.65 & 87.29 & 9.51 & 91.35 \\
    4 & VGG V4 & mae & 3 & \textbf{116.08} & 3.6 & 80.28 & 19.72 & 96.4 & 24.57 & 80.28 \\
    \hline
    5 & VGG V1 & ccl & 3 & \textbf{125.88} & 5.24 & 87.96 & 12.04 & 94.76 & 19.72 & 87.96 \\
    6 & VGG V2 & ccl & 3 & 152.24 & 9.21 & 90.88 & 9.12 & 90.79 & 12.61 & 90.88 \\
    7 & VGG V3 & ccl & 3 & 163.95 & 10.85 & 91.46 & 8.54 & 89.15 & 10.97 & 91.46 \\
    8 & VGG V4 & ccl & 3 & \textbf{122.79} & 6.02 & 83.94 & 16.06 & 93.98 & 16.93 & 83.94 \\
    \hline
    9 & VGG V4 & mae & 5 & \textbf{116.38} & 8.77 & 82.97 & 17.03 & 91.23 & 12.15 & 82.97 \\
    10 & VGG V4 & mae & 7 & 243.97 & 17.4 & 72.1 & 27.9 & 82.6 & 5.71 & 72.1 \\
    11 & VGG V4 & mae & 9 & 123.49 & 7.65 & 86.99 & 13.01 & 92.35 & 14.25 & 86.99 \\
    12 & VGG V4 & mae & 11 & 131.46 & 8.99 & 90.2 & 9.8 & 91.01 & 12.79 & 90.2 \\
    13 & VGG V4 & mae & 13 & 149.36 & 9.27 & 90.95 & 9.05 & 90.73 & 12.55 & 90.95 \\
    \hline
    - & Ensemble on imbalanced data & - & - & \textbf{118.46} & \textbf{5.38} & \textbf{89.29} & 10.71 & \textbf{94.62} & 19.53 & \textbf{89.29} \\
    \hline
    - & Ensemble on randomized balanced data & - & - & 118.47 & 5.38 & 89.29 & \textbf{10.70} & 94.62 & \textbf{94.31} & 89.29 \\
    \hline
    \end{tabular}
    \label{table:1}
    \end{sidewaystable}

    Concerning the localization subtask, VGG V1 and VGG V4 models (i.e., Models \#1 and \#4) performed best by achieving — on average — a lower Euclidean distance between predicted and actual TC center coordinates with respect to VGG V2 and VGG V3 (i.e., Models \#2 and \#3). This result still holds when the CCL is used as loss in the place of MAE (i.e., Models \#5 to \#8). All these models were trained with a kernel size of 3. 
    The VGG V4 model trained with a kernel size of 5 and the MAE loss (Model \#9) produced similar results to Model \#4, by achieving an Euclidean distance of 116.38 km. Moreover, increasing the kernel size up to 13 seems not to further improve the localization error. Nevertheless, increasing the kernel size allows better extracting spatial TC patterns and, thus, reducing the number of FN (i.e., Models \#12 and \#13). In general, as it can be observed from the results reported in Table \ref{table:1}, a higher localization error in terms of Euclidean distance corresponds to a reduction of the FN rate (lower is better).
    Regarding the classification subtask, FP and FN rates are in trade-off, thus a lower FN rate corresponds to a higher FP rate as outlined in the table. However, for studies like this addressing the detection of EWEs it is generally more appropriate having a higher number of FPs rather than FNs. Indeed, a false alarm (FP) is preferable, in such a case, against situations in which models incorrectly classify a TC as not present (FN).

    \subsection{Detection through the ML ensemble approach}\label{sec31}
\begin{sidewaysfigure}
        \noindent\includegraphics[width=0.6\textwidth]{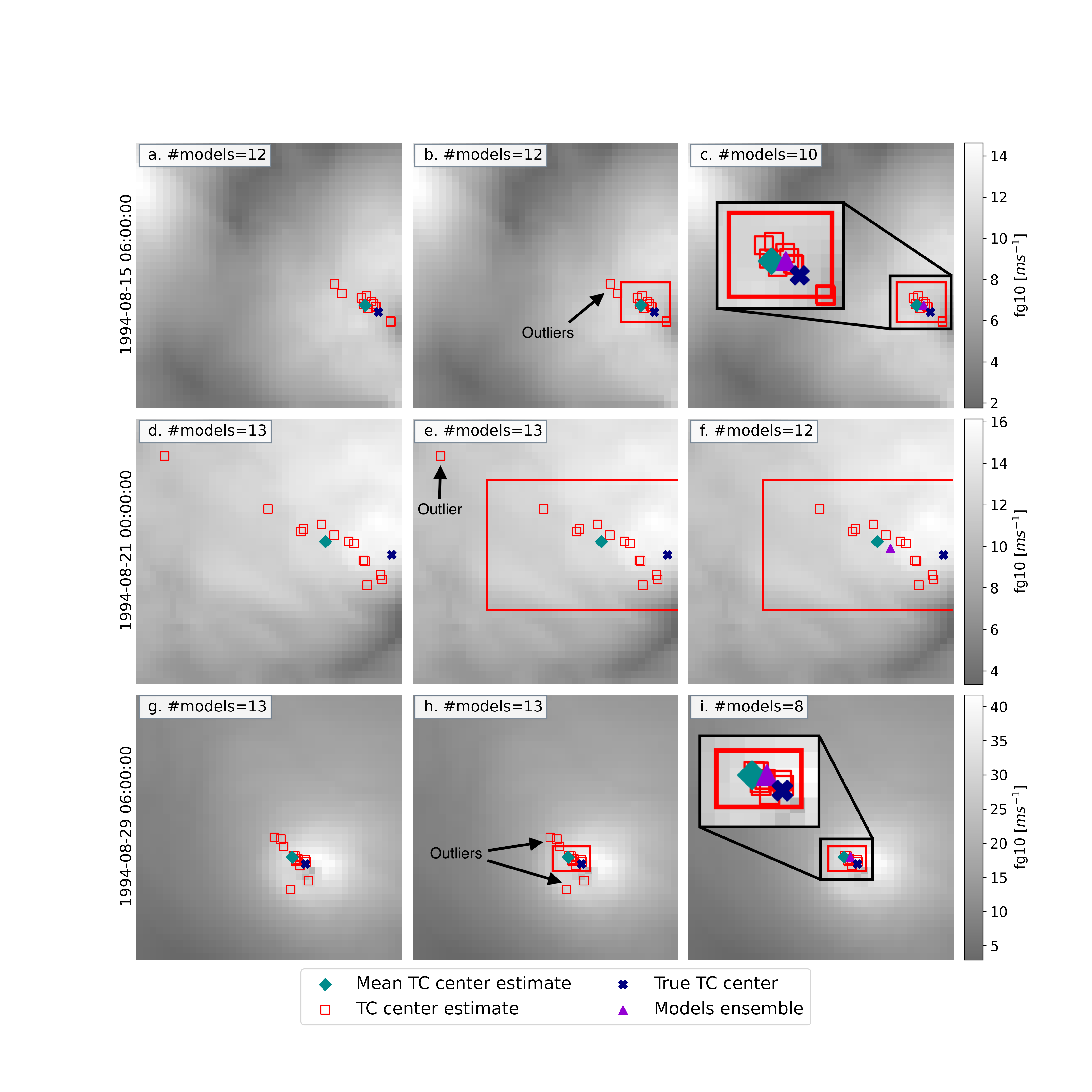}
        \caption{ML ensemble approach applied on three different time steps of TC John lifetime (rows), overlaid on the \textit{fg10} variable.  In each row, panels represent a particular stage of the proposed procedure. The number of models involved is reported in each panel and their TC center estimates are depicted as red squares, while the actual TC center is represented as a dark blue cross. Their average is reported as a green diamond. In center panels (b, e and h) the IQR is applied to detect outliers among models’ TC center estimates. In the right panels, the model ensemble average (purple triangle) is computed only considering inlier values.}
        \label{fig:ensemble_example}
    \end{sidewaysfigure}

    Figure \ref{fig:ensemble_example} shows the ML ensemble approach applied on three different patches during TC John evolution (August, 11\textsuperscript{th} to September, 13\textsuperscript{th} 1994), overlaid on the \textit{fg10} variable. Each row in Figure \ref{fig:ensemble_example} refers to a specific time step of John's lifetime, whereas each column describes a particular step of the proposed localization approach. For instance, Panel a) reports TC center estimates provided by 12 models (red squares) out of 13, thus only one model predicted the absence of a TC in the input patch. In this case, the minimum consensus of 7 is reached. In terms of localization error, the mean TC center estimate of such 12 models (green diamond) is 61.76 km far from the actual TC center (dark blue cross). Subsequently, in Panel b), the IQR method allows detecting 2 outliers (red squares outside the red box). Therefore, Panel c) reports only 10 remaining inliers (red squares inside the red box) along with the ensemble TC center estimate (purple triangle). Thanks to the IQR method the ensemble TC center estimate (purple triangle) is closer to the actual TC center (dark blue cross) than the initial mean estimate (green diamond). In particular, the distance between the actual TC center and the one provided by the IQR method applied on the ensemble is 39.06 km, with a localization improvement of about 37\%. The same description also holds for the examples reported in Panels d-i), where respectively 12 (in f)) and 8 (in i)) models contribute to the computation of the ensemble TC center estimate, with a localization improvement of about 21\% and 37\%, respectively. It is worth noting that the proposed procedure consisting in the combination of IQR and ensemble approach is of general value and it is particularly suited in such situations in which ML models predictions are spread. Panels d-f) in Figure \ref{fig:ensemble_example} represent an example of such behavior.  

    The aforementioned procedure was applied on the entire test set. The ensemble approach allows obtaining an average localization distance between the actual TC center and the estimated one of about 118.46 km (see the second to last row in Table \ref{table:1}). An average localization distance of 127.5 km would have been obtained if the IQR method was not applied. This results in an improved localization accuracy of 9 km. 
    From a classification perspective, evaluation of the ensemble on the test set reports 5.38\% of FPs and 10.71\% of FNs respectively (second to last row in Table \ref{table:1}). On one hand the recall metric was 89.29\%, meaning that the models’ ensemble is highly confident in identifying TC centers. On the other hand, considering that the number of FPs (52,051) is more than 4 times greater than the number of TPs (12,634), the precision metric is not very high (i.e., 19.53\%). Since there is a strong imbalance between \textit{positive} and \textit{negative} samples, these results must be carefully interpreted as it may result in a misleading perception of the ensemble classification skills \cite{koehler1996}. To this extent, a further evaluation procedure was conducted over $10,000$ balanced subsets randomly sampled from the test set. Each subset is composed of $16,000$ test patches. Both the randomness and the high number of subsets guarantees the statistical meaningfulness of the experiment. The evaluation metrics (reported in Section \ref{sec234}) were computed on each of the $10,000$ subsets and averaged together, and the results are reported in Table \ref{table:1} (last row).
    It can be noted that besides the precision metric, which showed a substantial increment to 94.31\%, the others were not affected at all. In summary, the ensemble shows good classification skills from both precision and recall standpoints on such a balanced dataset. Therefore, it further proves that the ensemble approach is more accurate than using a single ML model for TC detection.

    \subsection{Sensitivity Analysis of the ML ensemble approach}\label{sec32}
\begin{sidewaysfigure}
        \noindent\includegraphics[width=\textwidth]{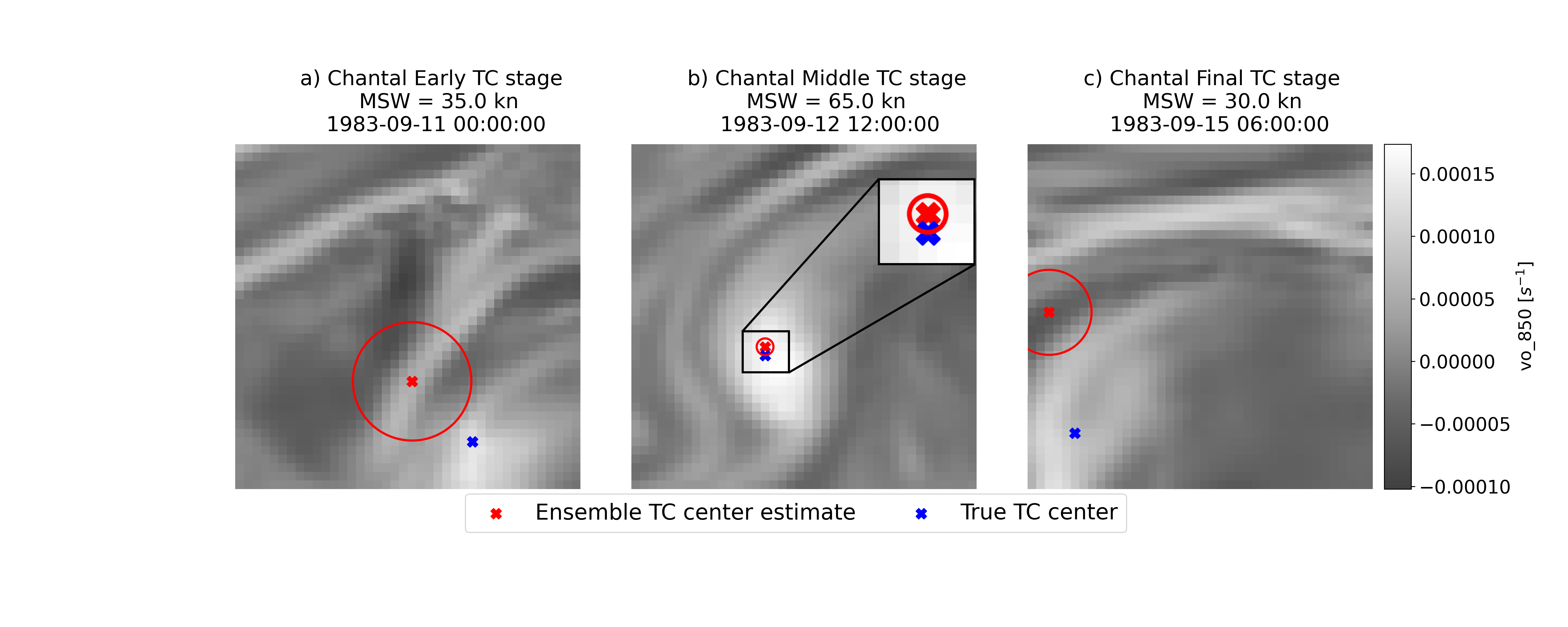}
        \caption{TC Chantal \textit{vo\_850} spatial patterns during early (left panel), middle (middle panel) and final (right panel) stages of its lifecycle. The ensemble TC center estimate (red cross) along with the true one (blue cross) is represented. The standard deviation of the ensemble TC center estimate is represented through the red circle.}
        \label{fig:chantal_evolution}
    \end{sidewaysfigure}

    A sensitivity analysis of the proposed approach was applied to assess the performances of the ML ensemble approach during various phases of the TC lifecycle that are typically characterized by different intensities of Maximum Sustained Wind (MSW), as registered in IBTrACS. As an example, in Figure \ref{fig:chantal_evolution} models ensemble predictions are reported for the TC Chantal (10–15 September 1983), overlayed on the \textit{vo\_850} input driver. In particular, three time steps over the Chantal lifecycle are shown, namely 1983-09-11 at 00.00, 1983-09-12 at 12.00 and 1983-09-15 at 06.00, respectively. In the early and final stages (i.e., a) and c) panels) the \textit{vo\_850} variable does not show the typical circular spatial patterns surrounding the storm eye and indeed the models ensemble struggle to estimate accurately the actual TC center (blue cross). This results in spread predictions and thus a higher standard deviation of the ML ensemble (red circle). To explain this situation, the MSW was retrieved from the IBTrACS dataset for the corresponding timesteps. The early and final stages of the Chantal TC are characterized by MSW speeds of 35 and 30 knots (i.e., weak TCs in IBTrACS), respectively. On the contrary, during the middle stage of its evolution (Panel b)), when the storm gains more strength (i.e., the MSW increases to 65 knots), spatial circular patterns of the \textit{vo\_850} become more evident around the eye. This makes the detection task easier and leads to a lower localization error between predicted and actual TC center location. Indeed, models involved in the ensemble predict approximately the same position, leading to a lower standard deviation (i.e., circle radius in Panel b) is smaller) from the ensemble TC center estimate.

    \subsection{Test Cases: Keoni and Julio Storms}
\begin{sidewaysfigure}
    \noindent\includegraphics[width=40pc]{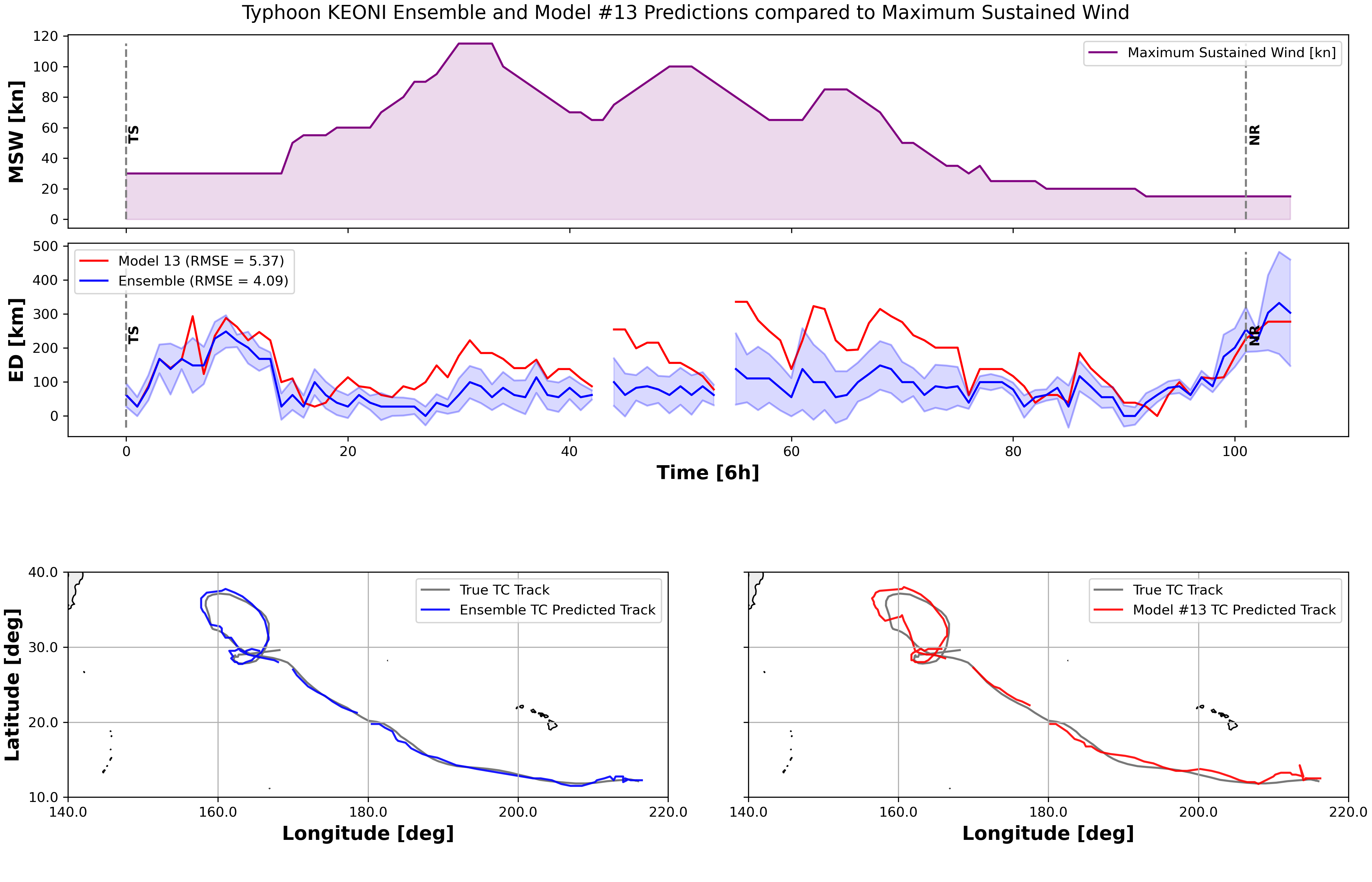}
    \caption{Comparison between performance of the ML ensemble against Model \#13 on TC Keoni track. The first panel depicts MSW during the cyclone evolution. The second one shows the Euclidean distance between real and predicted TC center estimates for both Model \#13 (red line) and the ensemble (blue line), with the standard deviation among models in agreement (light blue area) for the latter. Last two panels represent, respectively, the TC geographical coordinates predicted by the ensemble (blue line) and Model \#13 (red line) along with real Keoni tracks (gray line) over its lifetime. Discontinuities in middle and last panels time series correspond to time steps in which either Model \#13 or the ensemble did not detect the TC.}
    \label{fig:keonicomparison}
    \end{sidewaysfigure}

\begin{sidewaysfigure}
    \noindent\includegraphics[width=40pc]{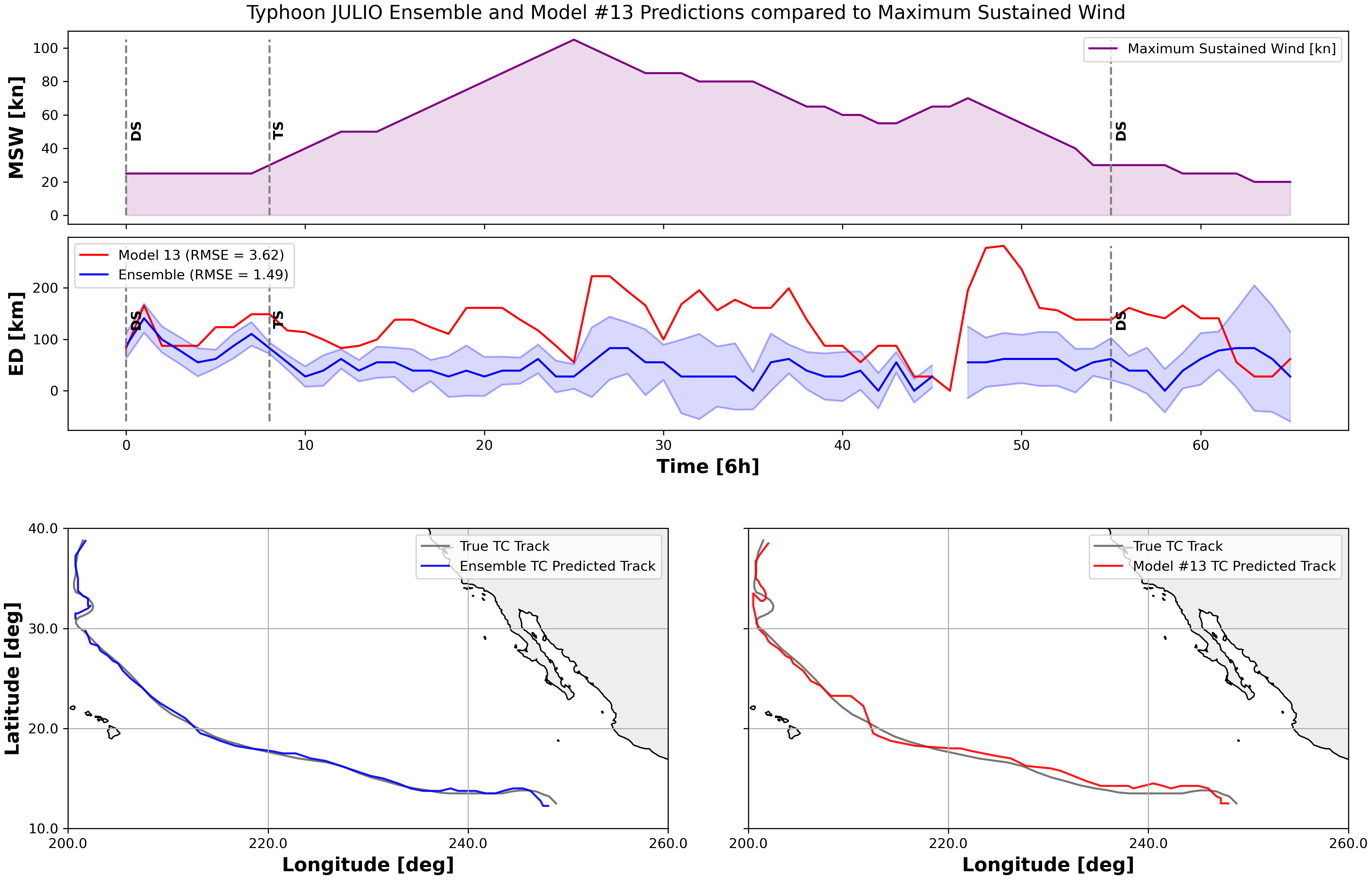}
    \caption{Comparison between performance of the ML ensemble against Model \#13 on TC Julio track. The first panel depicts MSW during the cyclone evolution. The second one shows the Euclidean distance between real and predicted TC center estimates for both Model \#13 (red line) and the ensemble (blue line), with the standard deviation among models in agreement (light blue area) for the latter. Last two panels represent, respectively, the TC geographical coordinates predicted by the ensemble (blue line) and Model \#13 (red line) along with real Julio tracks (green line) over its lifetime. Discontinuities in middle and last panels time series correspond to time steps in which either Model \#13 or the ensemble did not detect the TC.}
    \label{fig:juliocomparison}
    \end{sidewaysfigure}

    Model \#13 was used to evaluate the performance of the ensemble approach with respect to a single model. Model \#13 was selected since it showed more balanced localization and classification metrics (see Table \ref{table:1}). Specifically, the Keoni and Julio TCs that occurred in the domain of interest were analyzed. The centers detected by Model \#13 and the ensemble over the TCs evolution were reported in Figures \ref{fig:keonicomparison} and \ref{fig:juliocomparison}, respectively. Both figures are organized as follows: the upper panel represents the MSW of the TC, expressed in knots. Middle panel shows the Euclidean distance between true and predicted TC center coordinates, expressed in kilometers, produced by Model \#13 and by the ensemble approach. Lastly, lower left and lower right plots represent the true TC track (in gray) along with predictions from the ensemble (in blues) and Model \#13 (in red), respectively.

    \noindent\textbf{TC Keoni} \\
    The TC Keoni occurred from 9\textsuperscript{th} August to 4\textsuperscript{th} September 1993. During its lifecycle it became a hurricane characterized by strong winds that reached 115 knots of speed before starting to lose its intensity from 19\textsuperscript{th} August until early September.

    From Figure \ref{fig:keonicomparison} it can be evinced that early and final stages of Keoni lifecycle are characterized by low MSW, and therefore both Model \#13 and the ensemble provided TC center estimates with a higher Euclidean distance from the actual TC center coordinates. On the other hand, as the cyclone gains more strength, input drivers’ spatial features around the eye become more evident, thus making the TC detection easier for both the ensemble and Model \#13. As a result, Euclidean distance is lower than the other stages This analysis is in line with the one made in Section \ref{sec32} about the relationship between MSW and localization error.
    
    The time steps in which the TC was not detected (from now on referred to as discontinuities) are the same for both Model \#13 and the ensemble. This means that most of the models in the ensemble did not find a TC in the corresponding time step. Considering the overall TC trajectory, the ensemble better converges to the actual one provided by IBTrACS (lower left panel), as it implicitly exploits the contribution from all the models to provide more accurate TC center estimates. Conversely, Model \#13 (lower right panel) provides a higher localization error in TC center estimates even though it catches the overall trajectory pattern.

    \noindent\textbf{TC Julio} \\
    Similarly to the Keoni test case, the early and final stages of TC Julio are characterized by low MSW (upper panel) and higher localization errors for both the ensemble and Model \#13, as reported in Figure \ref{fig:juliocomparison}. Nevertheless, even though in the aforementioned stages the cyclone is classified as a Disturbance Storm, the models are still able to capture the phenomena in this status. This is a remarkable result since both Disturbance Storm and Not Reported TCs were filtered out from the training set and therefore their characteristics have not been shown during training. 
    
    Over the TC evolution, the Euclidean distance remains stable on average — especially for the ensemble — and slightly increases as the typhoon dissipates its energy. A discontinuity appears only for the model's ensemble, thus showing that Model \#13 detected the TC in the corresponding time step, but most of the other models failed (middle panel). Another interesting aspect worth highlighting is the large difference between the localization error of the ensemble model and the Model \#13 for the entire TC lifetime. This clearly demonstrates that the ensemble localization approach is more robust than a single model prediction, thus better converging to the real cyclone trajectory over its evolution, as depicted by lower left and right panels.

\section{Conclusion and Discussion}\label{sec4}
The present study proposed a TC detection approach based on ML that aims at identifying and localizing TCs’ center in terms of geographical coordinates by exploiting six climatic drivers that are related to the genesis and sustainment of such EWEs. 

Given the inherent complexity of the detection task, trusting the estimation of the TC center position through a single ML model would have led to unreliable results. Therefore, an ensemble approach was proposed to integrate the knowledge learnt by different ML models that are trained for the same detection task. The ensemble relies on 13 VGG-like architectures that are trained with distinct hyperparameters configurations on the same input-output pairs. This allows extracting different intrinsic patterns and features related to the TC evolution during its lifetime, as well as reducing the uncertainty associated with its center position estimate. The present approach is extendable either by adding new ML models to the ensemble or improving the current ones to get higher classification and localization performance.  

ERA5 reanalysis data, concerning six input climatic drivers, were jointly exploited with IBTrACS historical records to train and test the designed models. Reanalysis data combine model simulations with observations to provide the best state representation of different climatic variables in the past.

However, as recognized by \citeA{hodges2017}, no assimilation of TCs is performed in ERA5, unlike other reanalyses such as JRA-55 or NCEP-CFSR datasets. Nonetheless, the ensemble exhibits a  good accuracy, specifically providing 10.71\% of FN and 5.48\% of FP on the test set (Section \ref{sec31}). The IQR method applied to the ensemble as an outlier filtering step further improved the performance, thus leading to an average localization accuracy of 118.46 km (almost $1^{\circ}$ of latitude) on the test set. Indeed, \citeA{zarzycki2021} and \citeA{roberts2020} evaluated a series of metrics on ERA5 with comparable performance to JRA-55 and NCEP-CFSR: the main reason can be found in the enhanced resolution of ERA5 with respect to the previous ERA-Interim product. This motivated the use of ERA5 reanalysis for the six input climate drivers in this study. Moreover, the presented processing methodology of ERA5 maps led to non-cyclone-centric input patches, i.e., $40 \times 40$ images in which the TC center can occur in any position, not necessarily in its center. In this way, ML models were able to learn the drivers spatial patterns and characteristics related to the presence of the TC inside the patch, regardless of its position. As a result, beyond Tropical, Subtropical and Extra Tropical storms, the ensemble was capable of localizing also centers of NR (Not Reported) and DS (Disturbance Storms) cyclones with a low error, even though they were not included in the training set, thus demonstrating good generalization capabilities of the proposed approach. Furthermore, tiling ERA5 maps into non-overlapping patches of fixed size allowed ML models to detect multiple TCs that can simultaneously occur in the joint North Atlantic and Pacific geographical domain targeted in this study. The application of the proposed approach to other formation basins was not assessed in this study and will be subject to future investigation.

Concerning the limitations of the present research, it is important to take into account uncertainties related to IBTrACS and ERA5 data that may lead to biased TC center positioning. In particular, IBTrACS provides TC center geographical coordinates aligned on a $0.1^{\circ} \times 0.1^{\circ}$ grid and with an uncertainty that is inversely proportional to the storm intensity \cite{ibtracs_doc2019}. ERA5 maps, on the other hand, are provided on a $0.25^{\circ} \times 0.25^{\circ}$ grid. Therefore, TC centers were aligned on the ERA5 grid, as a preprocessing step (Section \ref{sec222}). As a result, all the sources of uncertainty implicitly affected both ML training and inference. As higher resolution reanalysis data will become available, the inherent uncertainty will also be reduced.

The proposed study focused only on analyzing the performance of the ML TC detection solution. As a future activity, a comparison with deterministic solutions like TStorm from National Oceanic and Atmospheric Administration (NOAA) (\url{https://www.gfdl.noaa.gov/tstorms/}) is envisioned to understand how the ML-based approach performs with respect to a deterministic tracking schema. 

Moreover, it  must be noted that the workflow for supporting the TC detection solution presented here is very complex and composed of heterogeneous data and software components. It requires large-scale data handling solutions, jointly with ML algorithms and access to High Performance Computing infrastructure. As next step the authors aim at developing an integrated pipeline that can apply the pre-processing and ML model pipeline directly to the output of an ESM simulation. This effort is currently ongoing in the frame of the eFlows4HPC EU project \cite{ejarque2022}. Finally, the ensemble approach proposed here will be explored in future work, in the context of the interTwin project (https://www.intertwin.eu/), by exploiting climate projection data from CMIP experiments with the aim of providing an indication on how climate change affects TCs frequencies and locations in the future.

\section*{Open Research Section}

The datasets used in this study are freely accessible from public repositories:
    \begin{itemize}
    \item Copernicus ERA5 reanalysis datasets:
    \begin{itemize}
        \item Single levels(i.e., mean sea level pressure, 10m wind gust since previous post-processing and  instantaneous 10m wind gust) : \url{https://cds.climate.copernicus.eu/cdsapp#!/dataset/reanalysis-era5-single-levels?tab=overview} (\cite{hersbach2023a})
        \item Pressure levels (i.e., relative vorticity at 850 mb, temperature at 300 mb and temperature at 500 mb): \url{https://cds.climate.copernicus.eu/cdsapp#!/dataset/reanalysis-era5-pressure-levels?tab=overview} (\cite{hersbach2023b})
    \end{itemize}
    \item International Best Track Archive for Climate Stewardship (IBTrACS) from National Centers for Environmental Information (NCEI): \url{https://www.ncei.noaa.gov/data/international-best-track-archive-for-climate-stewardship-ibtracs/v04r00/access/csv/} (\cite{knapp2010, knapp2018})
    \end{itemize}
    \begin{itemize}
    \item Source code for the Tropical Cyclones Detection approach presented in this work: \url{https://github.com/CMCC-Foundation/ml-tropical-cyclones-detection}
    \end{itemize}

\acknowledgments
This work was supported in part by the eFlows4HPC and InterTwin projects. eFlows4HPC  has received funding from the European High- Performance Computing Joint Undertaking (JU) under grant agreement No 955558. The JU receives support from the European Union’s Horizon 2020 research and innovation programme and Spain, Germany, France, Italy, Poland, Switzerland and Norway. In Italy, it has been preliminarily approved for complimentary funding by Ministero dello Sviluppo Economico (MiSE) (ref. project prop. 2659). InterTwin has received funding from Horizon Europe under grant agreement No 101058386. Moreover, the authors would like to thank dr.  Enrico Scoccimarro from the CSP (Climate Simulations and Prediction) Division of the CMCC for his scientific support.

%
%

\bibliography{main}

%
%
%
%
%

\end{document}